\documentstyle[12pt,epsf,fleqn]{elsart}

\def\figuredir{}
\def\psfiguredir{}

\def \ni{\noindent}

\def \znbb {$0\nu\beta\beta$}
\def \tnbb {$2\nu\beta\beta$}
\def \be {\begin{equation}}
\def \ee {\end{equation}}

\def\heimo{HEI\-DEL\-BERG--MOS\-COW--Ex\-pe\-ri\-ment}

\def\ex{experiment}

\begin{document}
\begin{frontmatter}
\title{A Large Scale Double Beta and Dark Matter Experiment: GENIUS}
\ni {\bf Authors} 

\author{J. Hellmig and H.V. Klapdor--Kleingrothaus}
\address{Max--Planck--Institut f\"ur Kernphysik, Heidelberg, Germany}

\begin{abstract}
The recent results from the HEIDELBERG--MOSCOW experiment have demonstrated
the large potential { of double beta decay} 
to search for new physics beyond the Standard Model.
To increase by a major step the present sensitivity for double beta decay and 
dark matter search much bigger
source strengths and much lower backgrounds are needed than used in 
experiments under operation at present or under construction. We { present}
here { a study of a project proposed recently \cite{kla97},} which would operate one
ton of 'naked' enriched GErmanium-detectors in liquid NItrogen as shielding
in an Underground Setup (GENIUS). 
It improves the sensitivity to neutrino masses to 0.01 eV. 
A ten ton version would probe neutrino masses even down 
to 10$^{-3}$ eV. The first version would allow to test the atmospheric
neutrino problem, the second at least part of the solar neutrino problem.
Both versions would allow in addition significant contributions
to testing  several classes of GUT models. These
are especially tests of R-parity breaking supersymmetry models, leptoquark masses and
mechanism and right-handed W-boson masses comparable to LHC.
The second issue of the experiment is the search for dark matter in the
universe. The { entire} MSSM parameter space for prediction of neutralinos as dark
matter particles could be covered already in a first step of the full
experiment { - with the same purity requirements but}
using only 100 kg of $^{76}$Ge or even of natural Ge 
{ - } making the experiment competitive to LHC 
in the search for supersymmetry.

The layout of the proposed experiment is discussed 
and the shielding and purity requirements are studied using GEANT Monte Carlo
simulations. As a demonstration of the feasibility of the experiment first
results of operating a 'naked' Ge detector in liquid nitrogen are presented.
\end{abstract}
\end{frontmatter}
\newpage
\section{Introduction}
Searches for rare events, nuclear double beta decay \cite{kla97,kla9596},
and nuclear recoils from elastically scattered weak interacting massive
particles (WIMPS) { \cite{smi90,ber95,jun96,lew97,bec93,bed94a,bed94b,bed97b}} are performed to discover 
new particles and test new particle physics theories \cite{kla95c,zub97b}. This type
of experiments is in contest to high energy accelerator experiments in
the investigation of physics at very high energies. Two topics are of
greatest interest in high energy physics and in astrophysics. 
Test of the existence of a
so called SUperSYmmetry (SUSY) is one, if not the, major aim of the
Large Hadron Collider (LHC) \cite{ber96,bag96,bae97}, which will dominate the
high energy physics research in the next decade. Second, dark
matter, which manifests itself by its gravitational force, puzzles
astrophysicists since a long time \cite{smi90,ber95,jun96,lew97}.
A very close connection between both issues in addition to SUSY which could be
responsible for the cold dark matter (neutralinos), could be
established by non zero neutrino masses  as candidates for hot dark
matter, 
and especially degenerate neutrino mass scenarios 
\cite{raf96,petc94,ioa94,fri96,moh95,lee94,cal95b} 
can explain the recent observations by the COBE satellite for 
dark matter \cite{cobe92,pri96}.
Our new GErmanium in NItrogen Underground Setup (GENIUS), first 
proposed and presented by \cite{kla97},
is an experiment
which is optimized to address both issues. 
GENIUS is a large step beyond the
HEIDELBERG--MOSCOW--Experiment \cite{hei97a},
which is the most sensitive existing double beta decay experiment at present 
and for the next years and which has also given the most stringent limits on
WIMPS for several years \cite{bec93}. 
The GENIUS project,
which we describe in this paper, would allow a large step 
forward in sensitivity and could represent the future of this field. It would 
not only be unique in probing {\it absolutely} the neutrino mass down to 10$^{-2}$ 
or 10$^{-3}$ eV (all running neutrino oscillation experiments probe only 
differences of masses!) but could also decisively contribute to the solution 
of the atmospheric and solar neutrino problems. It is further expected to give
contributions to the research into SUSY, leptoquarks and right--handed 
W--boson masses comparable to the potential of LHC or NLC (for a 
detailed discussion { of the physics potential of GENIUS} see \cite{hk97}). 
Concerning dark matter it will be able
to cover the whole SUSY (MSSM) parameter space for predictions of neutralinos
as dark matter and this would be competitive to LHC concerning the search for 
supersymmetry. The costs of the experiment would be
a minor fraction of detectors prepared for LHC physics as CMS or ATLAS.
 
\subsection{$\beta\beta$-Decay Searches}
Nuclear double beta decay can be observed, if the single
beta decay is either energetically forbidden or through high angular momentum
differences between final and initial nucleus suppressed. The decay is
divided into two major modes, according to the number of emitted particles.

\begin{eqnarray} \label{g1}
& 2\nu\beta\beta & ^Z_AX \quad\rightarrow\quad ^{Z+2}_AX + 2 e^- + 2 \bar \nu_e 
\\
& 0\nu\beta\beta & ^Z_AX \quad\rightarrow\quad ^{Z+2}_AX + 2 e^-
\end{eqnarray}

The two neutrino \tnbb{} mode occurs under emission of two electrons
and two antineutrinos, whereas in case of the neutrinoless
\znbb{} mode only two electrons
are emitted. The \tnbb{} mode is allowed in the standard model of
particle physics and is already observed for 10 isotopes. The by far
more interesting decay mode is the \znbb{} mode, whose observation
implies the existence of new physics beyond the standard model. For 
possible mechanisms, aside from the Majorana neutrino exchange,
leading to this decay see \cite{kla97,kla9596,hir96b,hir96c,hir96,hir97plb,hir96ph}. 

\begin{figure}
\epsfxsize13cm
\epsffile{\figuredir 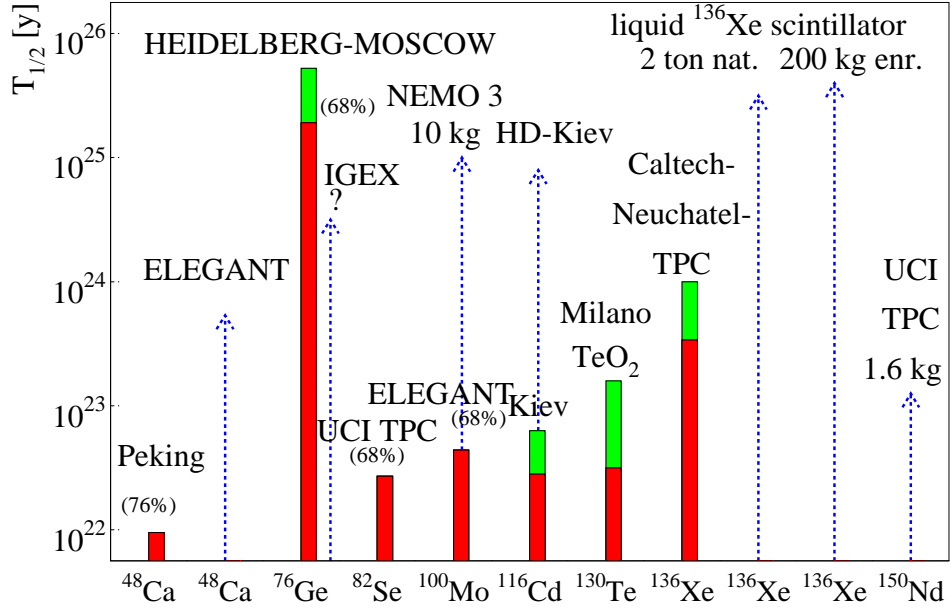}
\caption{Present half life limits (filled bars), 1997, and
`safe' expectations (light shaded bars) for the near future
(until the year 2000) and long term planned or 
hypothetical experiments (dashed lines) for the further future of 
the most promising $\beta\beta$-experiments.}\label{fstatus}
\end{figure}

\begin{figure}
\epsfxsize13cm
\epsffile{\figuredir 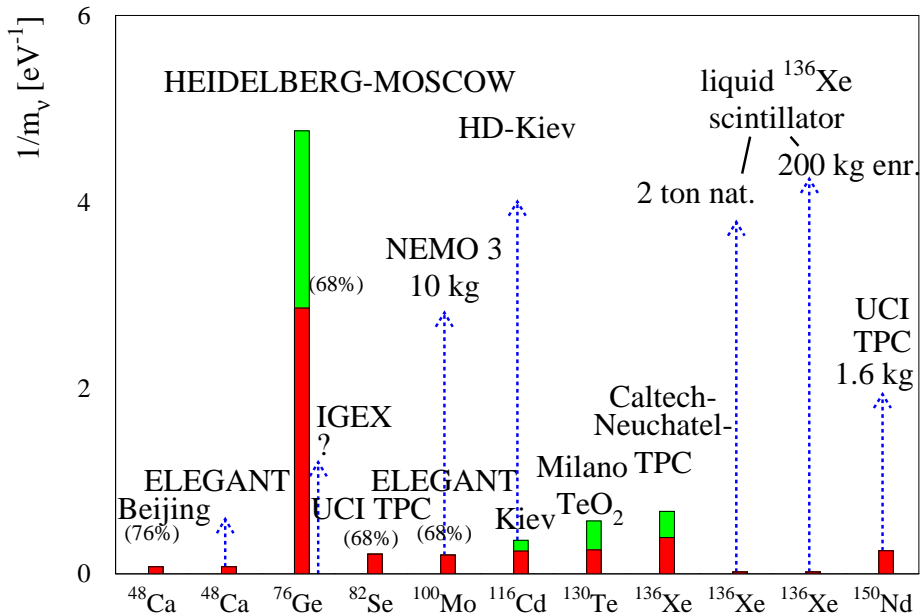}
\caption{Inverse neutrino mass limits from the neutrinoless double
decay half life limits of figure \ref{fstatus}}\label{fstatus2}
\end{figure}
The best presently existing limits for \znbb --decay half--lives and the 
deduced neutrino masses are plotted in figure \ref{fstatus} and \ref{fstatus2}.
$^{48}$Ca  \cite{ke91}, 
$^{82}$Se  \cite{ell92}, 
$^{100}$Mo  \cite{eji96}, 
$^{116}$Cd  \cite{dan95}, 
$^{130}$Te  \cite{ale94b}, 
$^{136}$Xe  \cite{vui93} and
$^{150}$Nd  \cite{moe95}. Other double beta decay setups as NEMO  \cite{nemo96}, 
the Gotthard $^{136}$Xe TPC experiment  \cite{joe94}, 
the $^{130}$Te cryogenic experiment  \cite{ale94b},
a new ELEGANT with $^{48}$Ca  \cite{kum95},
 etc., will not reach or exceed the $^{76}$Ge limit. 
For example a limit of 0.3 eV - almost the limit reached now by the
\heimo - is scheduled for the NEMO experiment for the
year 2005. The $^{48}$Ca experiment of the Osaka group plans to reach a limit
of 0.8 eV \cite{eji95}, just to mention the two most promising experiments 
behind the \heimo, which aims to a limit of 0.1 - 0.2 eV in the year 2000. 
So 0.1 eV is the definite limit for all presently existing or 
planned $\beta\beta$ experiments { (see also \cite{nor97})}.
Therefore the GENIUS experiment would be 
a step into a totally new half life and neutrino mass region, 
which is not accessible by any other double beta decay experiment.

\subsection{Direct Dark Matter Detection}
Weakly interacting massive particles (WIMPs) are major candidates for
the cold dark matter in the universe. Assuming a local WIMP halo density
of 0.3 - 0.7 GeV/cm$^{3}$ and a  mean WIMP velocity of 10$^{-3}$c
nuclear recoils below 100 keV from GeV/c$^{2}$ particles are expected.

\begin{figure}[ht]
\vspace*{-12cm}
\hspace*{-1cm}
\epsfysize=130mm\centerline{\epsfbox{\psfiguredir }}
\vspace*{10cm}
\caption{WIMP--nucleon cross section
limits (hatched region: excluded by the \heimo \protect{\cite{bec93}}
and the UKDMC NaI \ex{} \protect{\cite{smi96}};
solid line: DAMA result for NaI, see
\protect{ \cite{ber97,bot97}}) in pb for scalar interactions as
function of the WIMP--mass in 
GeV and of possible results from upcoming \ex{}s (dashed lines for
HDMS \protect{ \cite{bau97}}, CDMS \protect{ \cite{bar96}} and GENIUS). These 
\ex{}al limits are compared to
expectations (scatter plot) for WIMP--neutralino cross sections calculated in the
MSSM framework with non universal scalar mass unification
\protect{ \cite{bed97a}}.  The 90~\% allowed region \cite{ber97b}
(light filled area), which is further
restricted by indirect dark matter searches \cite{bot97b} (dark filled
area), could be easily tested with a 100 kg version of the GENIUS experiment.}
\label{zf}
\end{figure}

The favorite WIMP candidate is the lightest supersymmetric particle
(LSP) in the minimal supersymmetric standard model (MSSM).
The expected detection rates for this particle
are below 1 per day and kg detector mass
\cite{bed97a,bed94a,bed94b,bed97b,jun96},
which cannot be reached by any experiment at present, due to
backgrounds from sources such as natural radioactivity, neutrons or
nuclear beta decay. Therefore only limits on the WIMP--nucleon cross
section as a function of the WIMP mass are deduced. Figure \ref{zf}
gives the best { experimental}
results { (limits)} on the scalar WIMP--nucleon { scattering} 
cross section, the expectations for
experiments under construction and { theoretical expectations
calculated in the MSSM over the whole SUSY parameter space
\cite{bed97a}}
The GENIUS
experiment will be the only one, which could seriously test the MSSM
predictions over the whole SUSY parameter space (see section \ref{chapdmlimits}).

\section{The GENIUS Experiment}
\begin{figure}
\epsfxsize13cm
\epsffile{\figuredir 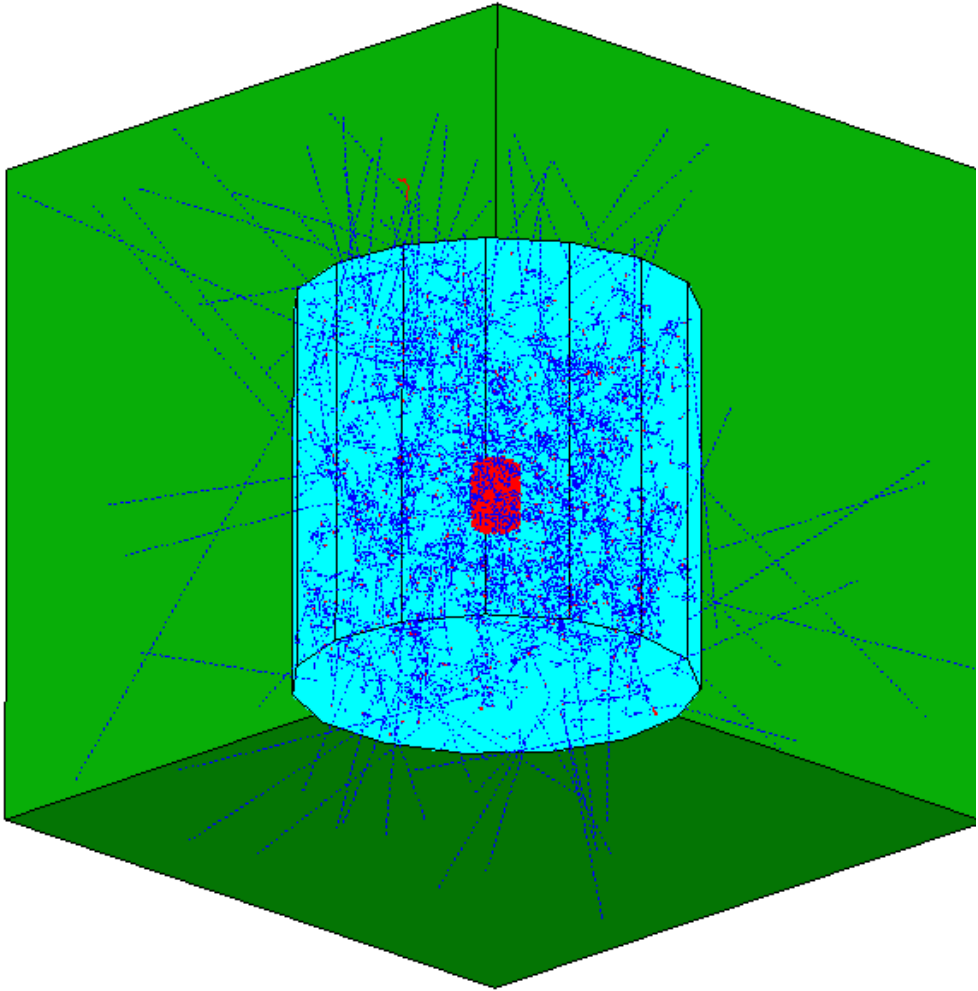}
\caption{Simplified model of the GENIUS experiment: 288 enriched
$^{76}$Ge detectors
with a total of one ton mass in the center of a 9 m high liquid nitrogen 
tank with 9 m diameter; GEANT Monte Carlo simulation of 1000 2.6 MeV 
photons { randomly distributed} in the nitrogen is also shown.}\label{fgenius}
\end{figure}

The idea of the GENIUS project is to operate { a large amount of}
'naked' enriched Ge--detectors
in a liquid shielding of very low--level radioactive material.
From the \heimo{} we learned that the main contributions to the
background spectrum arise from the cryostate system and the lead 
shielding. A substantial progress by using cleaner materials cannot be 
expected, without reducing the background in the materials next to 
the detectors. The 
possibility to measure at much lower radioactivity levels is 
demonstrated by the new solar neutrino experiments Super 
Kamiokande\cite{sup96}, SNO\cite{sno96} and Borexino\cite{bor96}. 
A very promising way  for operation of HPGe-detectors would be to put them
into a liquid, which cools them to their working temperature and shields
them from external radioactivity (see Fig. \ref{fgenius}). 
The idea to operate
germanium detectors in liquid nitrogen was earlier considered by
\cite{heusser}.

\subsection{The Shielding}
Table \ref{tabliquidprop} gives some
properties of nitrogen, argon and xenon.
All three liquids can be processed to very high purity.

\begin{table}[h]
\begin{minipage}{\textwidth}
\setcounter{mpfootnote}{0}
{\bf
\caption{Properties of the liquids nitrogen, argon and xenon.}\label{tabliquidprop}}
\vskip0.3cm
\centering
\begin{tabular}{lccc}
\hline
Liquid & Melting point [K] &Boiling point [K] & Density [g/cm$^{3}$]
\footnote{Density at normal pressure and boiling temperature, except
argon density at 81.7~K.}\\
\hline
nitrogen & 63 & 77 & 0.80\\
argon & 83 & 87 & 1.63\\
xenon & 161 & 165 & 3.52\\
\hline
\end{tabular}
\end{minipage}
\end{table}

The advantages of nitrogen are its low price and low temperature. The
drawback of a liquid nitrogen shielding is its low density, which
requires a very large setup to achieve sufficient shielding from
external activities.
Liquid argon shields better by a factor of two,
but contains the $\beta$--emitter $^{39}$Ar with 279 years half life,
which is produced through $^{38}$Ar(n,$\gamma$). The Q--value of 565 keV
is well below the Q--value of the $^{76}$Ge double beta decay, but the
electrons would contribute to the background in the dark matter recoil
energy region. Liquid xenon has a higher density, but its temperature is at 
the upper edge for the operation of HPGe-detectors and the price is an
additional problem. Altogether, the best choice for a liquid shielding would be
nitrogen.

A nitrogen tank could be designed in two approaches to achieve
lowest background. One possibility would be to use selected low
activity materials for the tank wall and an additional shielding
against radioactivity from outside. The second possibility would be to
use a standard design made of standard materials for dewar
production. In this case the diameter of the tank has to be big enough
to shield all activities from outside and from the tank walls. The
advantages of the first approach are a smaller detector size with
reduced material usage. But the
necessity to use selected materials and additional shieldings would
probably lead to higher costs, than in case of the second approach.
In the following we concentrate on the second approach.

\subsection{The Germanium Detectors inside the liquid shielding}
\begin{figure}
\epsfxsize13cm
\epsffile{\figuredir 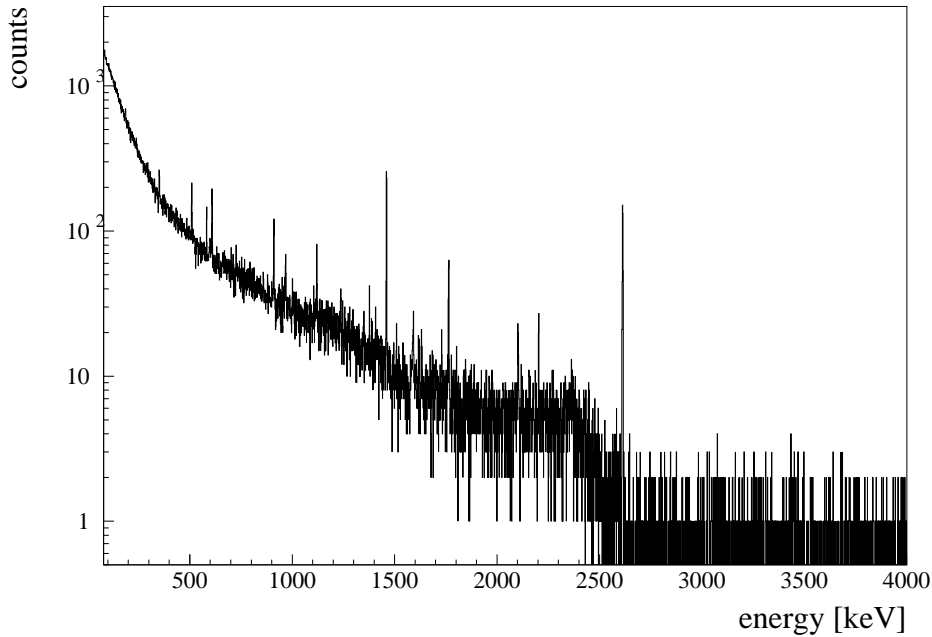}
\caption{Spectrum of a HPGe--detector operated in liquid nitrogen {
in the Heidelberg underground site}; lines of 
natural radioactivity and muon induced background above 3 MeV can be seen.}\label{fbg}
\end{figure}
\begin{figure}
\parbox[t]{7cm}{\epsfxsize6cm\epsffile{\figuredir 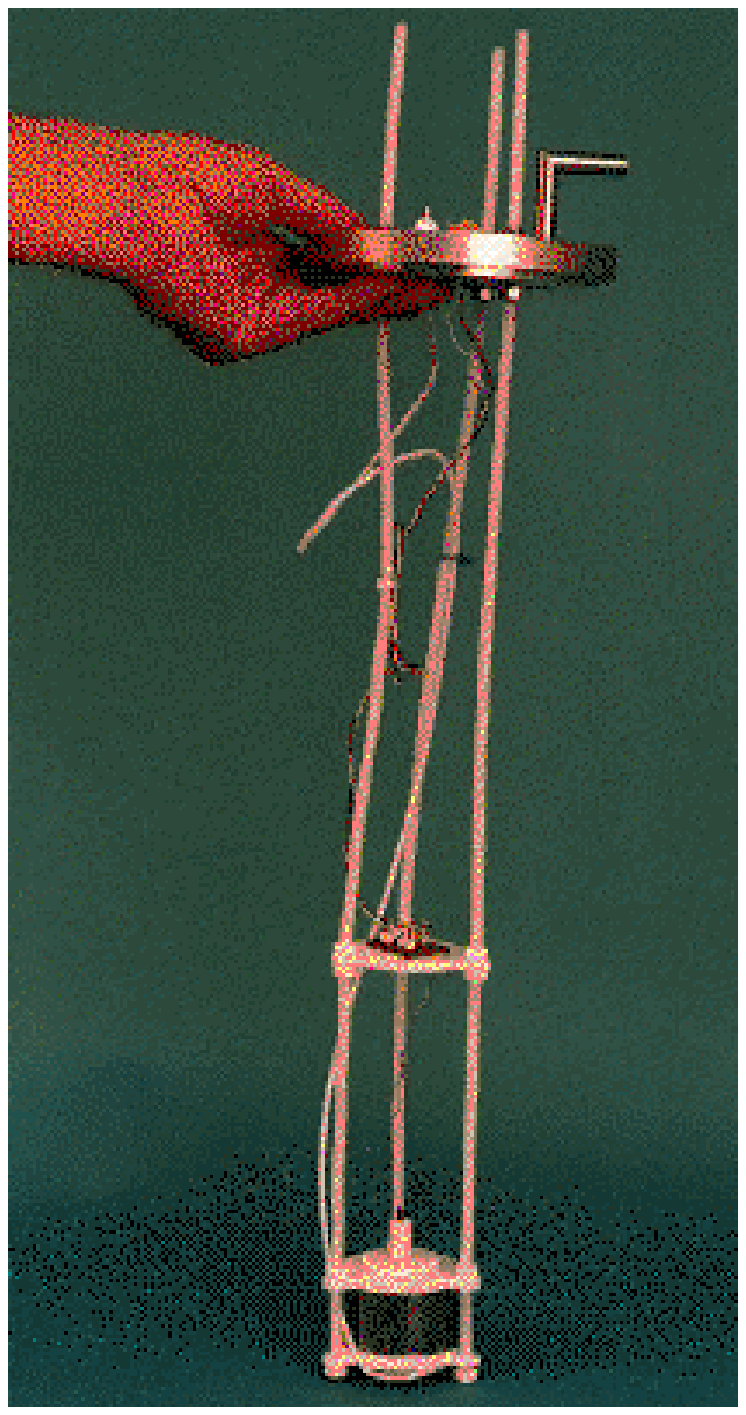}}
\parbox[t]{8cm}{\epsfxsize7cm\epsffile{\figuredir 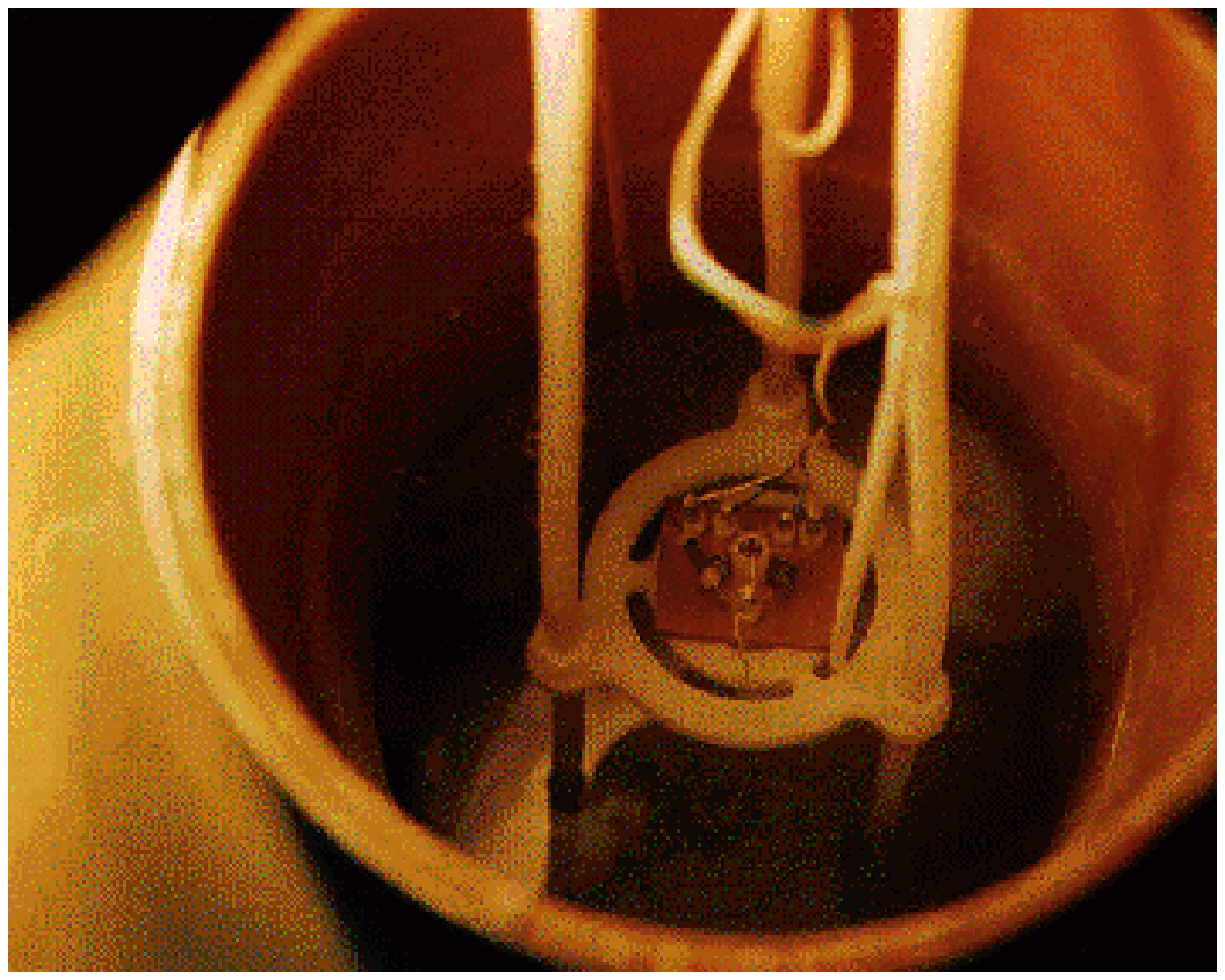}}
\caption{Setup for operation of one 'naked' Germanium detector in
liquid nitrogen: frame with FET platform and crystal (left) and view into
the liquid nitrogen dewar (right).}\label{fphoto}
\end{figure}

To demonstrate the possibility to operate Germanium detectors 
inside liquid nitrogen we used a p--type HPGe-detector. Figure \ref{fbg}
shows the spectrum measured with the detector inside a 50 l dewar. The dewar 
was surrounded by 10 cm lead. The FET was mounted on a small board inside the 
nitrogen 6 cm above the crystal and connected with 1 m long HV and signal 
cables to the preamplifier  (see figure \ref{fphoto}). 

\begin{figure}
\epsfxsize13cm
\epsffile{\figuredir 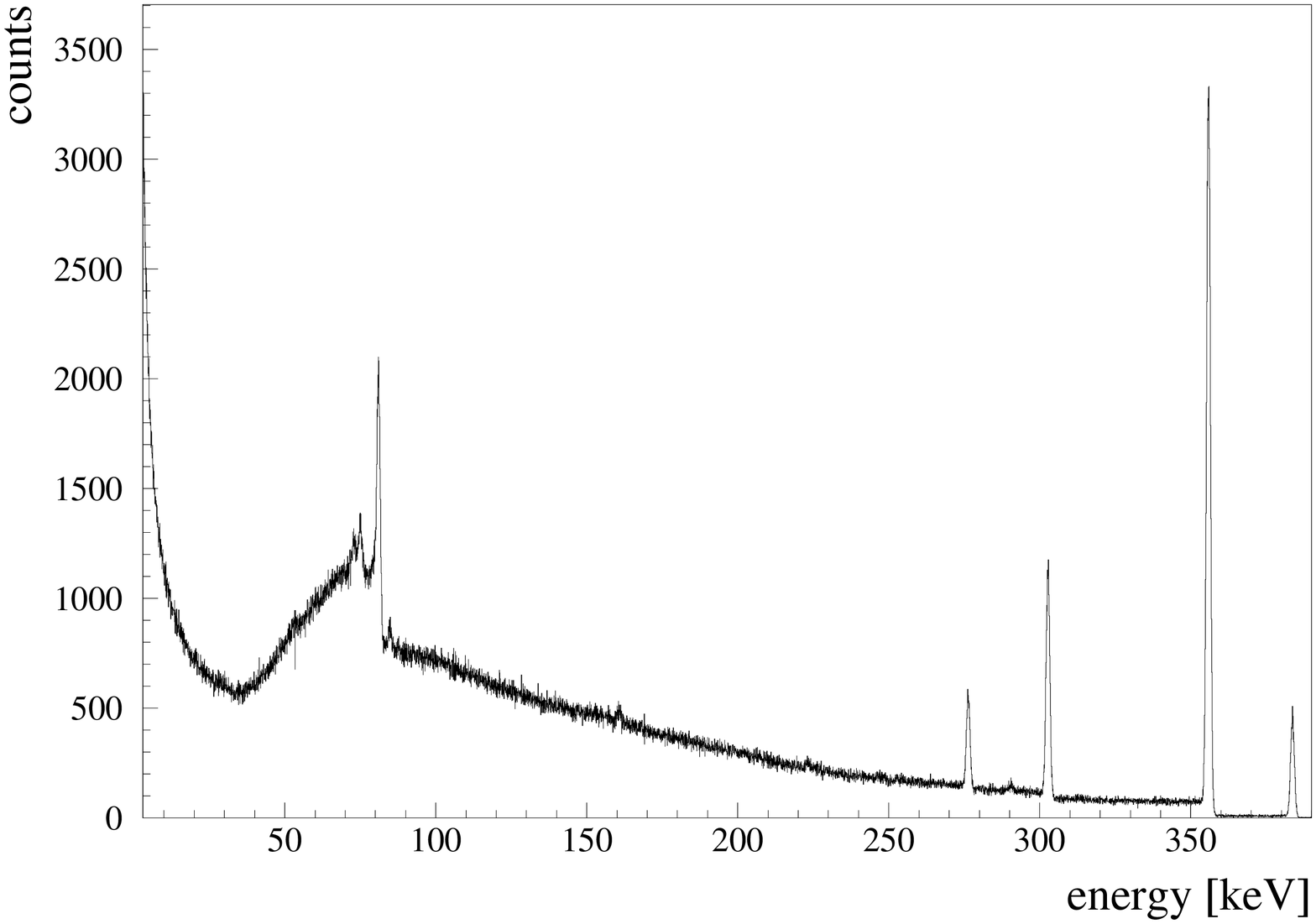}
\caption{Spectrum of a $^{133}$Ba source measured with a HPGe--detector 
operated in liquid nitrogen { in the Heidelberg underground
low--level laboratory}. X--rays from lead can be seen with energies
of 72.8~keV and 75.0~keV.}\label{fba}
\end{figure}

The energy resolution was measured with a $^{133}$Ba source 
at 81.0 keV to be 1.21 $\pm$ 0.01 keV and at 356~keV to be 1.51 $\pm$ 0.01 
keV. 
With a $^{60}$Co source 1.95~keV were measured at 1332~keV. 
The resolution is only about 
0.1~keV worse than in a standard cryostat, most probably due to a non 
polished gold foil used as signal contact. Note, that the TPI, a measure for 
the leakage current, changed only by one~\% between HV off and 4000 V.
Figure \ref{fba} shows the spectrum measured with a $^{133}$Ba
source. The lead X-rays and an energy threshold below 10 keV can also be seen.

A point, which should receive some attention for the GENIUS experiment,
might be electrical interference of several detectors, which has to be tested.
From the point of view of lowest
radioactive level the production and transportation of the detectors at
the Earth's surface have to be reconsidered. Already in the \heimo{}
lines from cosmogenic activities {\it inside} the detectors could be
identified. Production of the detectors inside the underground
laboratory is probably the most straightforward way to avoid 
the cosmogenic activation at the level of background aimed at. 

Muon induced background contributes about 10 \% to the spectrum in the
2 MeV region of the \heimo. This would mean, however, a large contribution to
the background in the GENIUS experiment. This background could be
suppressed by using an active shielding either enclosing the nitrogen
tank or by using scintillation light from the nitrogen itself. Since
pure nitrogen does not scintillate it has to be doped with some
substance. However, our simulations show, that an anticoincidence of the
288 detectors reduces the count rate of muon induced events sufficiently.

Neutron induced background has been proved to exist in the \heimo{} by
subtracting data accumulated with neutron shield from the data
accumulated without neutron shielding. However, this background cannot be
compared directly to that of the nitrogen environment in the GENIUS 
experiment leading to different effects to those from the lead and copper 
shielding. The low mass of nitrogen would thermalize the neutrons. 
Additionally there is for example the prominent $^{14}$N(n,p)$^{14}$C 
reaction which would contribute ($^{14}$C is a $\beta^{-}$--emitter with 
Q=156.5 keV) \cite{led87} and for the application of GENIUS for low energy 
measurements like WIMP direct detection one has to take care of intrinsic low
energy background irrelevant for double beta decay measurements.

The requirements for the data acquisition are:
\begin{itemize}
\item high energy resolution in the energy range from 10 keV to 2 MeV
\item event by event data acquisition
\item separated energy information of each detector
\end{itemize}
Calibration of the experiment could be done by introducing sealed
sources into the nitrogen tank. Positioning of one or more sources in
the middle of the germanium detectors would allow to get a good
energy calibration for the experiment. A calibration in the low energy
range would be particularly easy, since the unshielded detectors {
would be sensitive to} very low energy gamma rays. 

\subsection{Expected Performance}
To determine the size of the experiment and the required
purity levels we used the Monte Carlo code GEANT extended
for the simulation of radioactive decays. This version was already
successfully employed in the measurement of the \tnbb{} half life and
the investigation of background in the \heimo \cite{hei97a}. 

The { sources of} radioactive background in the
GENIUS experiment { can be differentiated}  according to their origin. The external
backgrounds arise from $\gamma$--fluxes and neutrons from the natural 
activities in the surrounding rock. The muon induced background is not
negligible in spite of six orders of magnitude reduction through the Gran
Sasso mountain.
Internal backgrounds are expected from impurities in
the vessel, the liquid nitrogen and the crystals themselves.

\subsubsection{Signal Rates}
The signal of the 1.77$\cdot10^{21}$y half life \tnbb{} decay is the dominating
feature of the GENIUS energy spectrum (see figure \ref{fbathsim}). 
After one year of measurement the \tnbb{} spectrum contains 4$\cdot10^{6}$ 
events. But only 0.3 events per year are expected from the neutrinoless
double beta decay, assuming the neutrino to have a mass of
10$^{-2}$eV, which corresponds to the ultimate sensitivity of the 1
ton experiment. The expected neutralino rates range from 10 to 
10$^{5}$~counts/keV$\cdot$t$\cdot$y. The design goal for the detector is to 
have the sensitivity to test 0.3 events in
the 2 MeV and 10 events in the 30 keV region. Since in the
low energy region the simulated \tnbb{}
spectrum contains by a factor of 10 more events, the \tnbb{} spectrum has 
to be subtracted. { Another promising way would be to exploit the
seasonal modulation of the dark matter flux \cite{ram97}}. Note, that
{ the \znbb{}} measurement is not affected by the \tnbb{} signal (which is a
major source of background in other experiments such as e. g. NEMO). This is a result
of the very good energy resolution of the Ge detectors.

\subsubsection{Estimate of Background}
The background { is} estimated using a simplified model of the GENIUS
experiment. It consists of 288 Ge--detectors of 3.6 kg each, arranged in
six radial symmetric layers of 48
detectors. The detectors occupy 1 m height and 1 m
in diameter in the center of a tank, which is 9 m in height and 9 m in
diameter and filled with
liquid nitrogen. The vessel is made of 2 cm thick steel. The
setup can be seen  in figure \ref{fgenius} together with a simulation of 1000
2615 keV photons randomly distributed inside the nitrogen.

\subsubsection{Activities in the nitrogen:}

\begin{table}
\begin{minipage}{\textwidth}
\setcounter{mpfootnote}{0}
{\bf
\caption{Required purity levels for the liquid nitrogen.}\label{trlevel}}
\vskip0.3cm
\centering
\begin{tabular}{lcc}
\hline
Isotope & Activity & Decays/year\\
\hline
$^{222}$Rn              & 0.05 mBq/m$^{3}$      & 8$\cdot 10^{5}$\\
$^{238}$U (4n+2 series) & 1$\cdot 10^{-15}$ g/g& 2$\cdot10^{5}$\\
$^{232}$Th (4n series)  & 5$\cdot 10^{-15}$ g/g& 3$\cdot10^{5}$\\
$^{40}$K                & 1$\cdot 10^{-15}$ g/g& 4$\cdot10^{6}$\\
\hline
\end{tabular}
\end{minipage}
\end{table}

\begin{figure}
\epsfxsize13cm
\epsffile{\figuredir 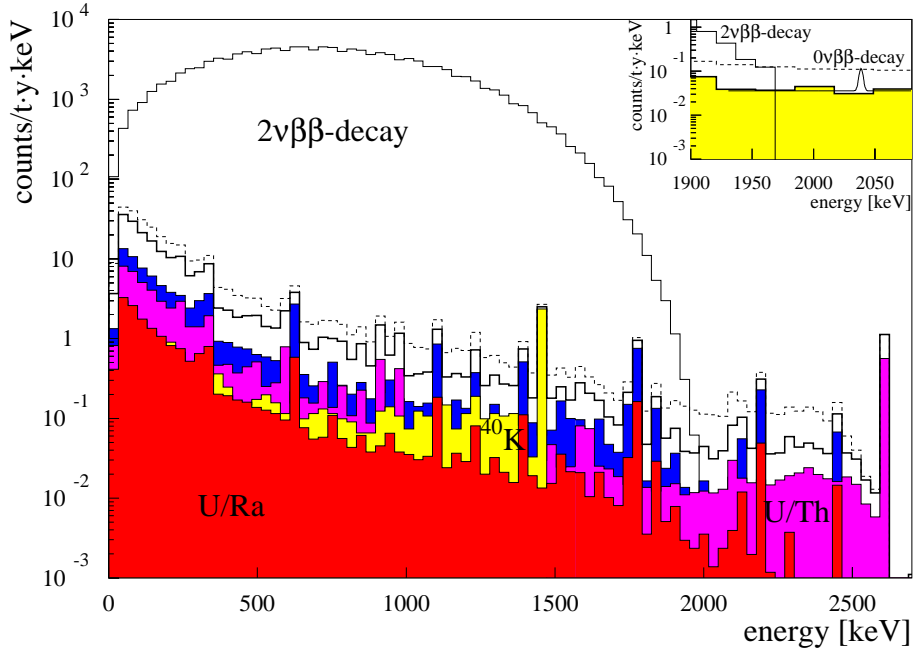}
\caption{Monte Carlo simulation of U/Ra, U/Th and $^{40}$K
(shaded), $^{222}$Rn (black histogram) activities in the liquid
nitrogen; the sum of the activities is shown with anticoincidence
between the 288 detectors (thick line) and without (dashed line); the
\tnbb -decay dominates the spectrum with 4 million events per year;
the impurity levels  are assumed as given in table
\ref{trlevel}.}\label{fbathsim}
\end{figure}

The main contributions of radioactivities inside the nitrogen are expected from
the nuclear decay chains of U/Ra with decays of $^{234}$Th, $^{234}$U,
$^{230}$Th, $^{226}$Ra,
$^{214}$Pb, $^{214}$Bi and $^{210}$Pb, U/Th with decays of $^{228}$Ac,
$^{228}$Th, $^{224}$Ra, $^{212}$Pb, $^{212}$Bi and $^{208}$Tl,
primordial $^{40}$K and $^{222}$Rn. To study
the required purity levels we simulated 28 million decays of each isotope
randomly distributed inside the nitrogen.  { The spectra of each isotope
from the decay chains are summed under the assumption that the chains are in equilibrium.}
The energy spectra of non--coincident events are shown in figure 
\ref{fbathsim} assuming purity levels as listed in table \ref{trlevel}.
All requirements except for the $^{222}$Rn contamination are less stringent,
than those which are already achieved in the { Counting Test
Facility (CTF)}
for the Borexino experiment.
The sum of all non--coincident events is plotted with a thick line and of 
all events with a dashed line in figure \ref{fbathsim}. 
The reduction through anticoincidence is about a factor of 10. The
counting rate in the region of interest for the neutrinoless double beta 
decay is  0.04 counts/keV$\cdot$y$\cdot$t. Below 100 keV the counting rate is 
about 10~counts/keV$\cdot$y$\cdot$t. 

\subsubsection{Activities in the vessel:}
\begin{figure}
\epsfysize9cm
\epsfxsize13cm
\epsffile{\figuredir 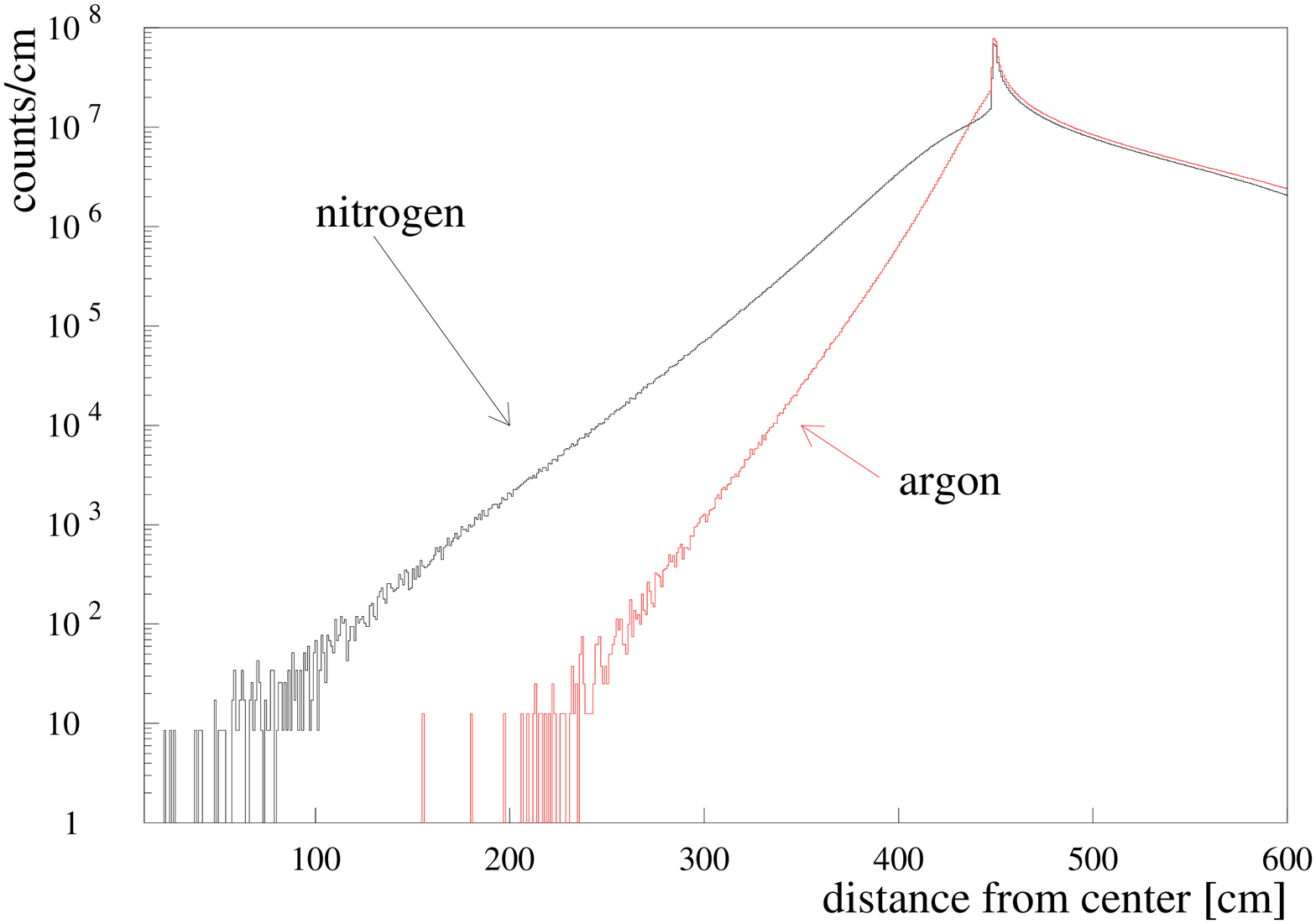}
\caption{{ Simulation of $^{208}$Tl activity randomly distributed in
the vessel; the histogram contains for each event the
distance of the nearest interaction to the detector center;
for comparison filling of liquid nitrogen and argon are simulated.}}\label{fig2614rad}
\end{figure}

The shielding of radioactivity from outside the nitrogen is shown
in figure \ref{fig2614rad}. It shows the distribution of the nearest
interaction to the center of the detector for 6$\cdot 10^{8}$ simulated
$^{208}$Tl decays { randomly distributed} inside the 2 cm thick
steel vessel surrounding the 
nitrogen. The radioactivity drops 0.8 orders of magnitude per meter
diameter. Liquid argon, which is shown as well, has a doubled
attenuation coefficient, because of the doubled density. 

From simulation of 1.1 billion $^{208}$Tl decays in the steel vessel only 8
deposit energy in a Germanium detector (see figures \ref{fig2614rad}
and \ref{fmuons}), due to
the shielding of the liquid nitrogen. This means,
an impurity concentration of 1$\cdot 10^{-8}$ g/g U/Th (40 mBq/kg) in
the steel vessel would 
contribute to the count rate in the Ge detectors as much as the
impurities in the nitrogen.

The shielding of radioactivity from outside the tank is not simulated
yet. But a first estimation can be made from the simulation of
impurities inside the steel vessel. Assuming that the 2 cm thick steel
(7.9 g/cm$^{3}$) vessel has no shielding effect the activity of 40
mBq/kg can be converted into a flux of 3.2$\cdot 10^{-4}$cm$^{-2}$s$^{-1}$,
which is by a factor of 50 lower than the flux of $^{208}$Tl 2.6 MeV photons 
of 1.5$\cdot 10^{-2}$cm$^{-2}$s$^{-1}$ measured in hall C of the Gran Sasso
lab. Therefore an additional shielding of for example 10 cm lead or
0.8 m liquid nitrogen has to be applied. 

\subsubsection{Muon induced background:}
\begin{figure}
\epsfxsize13cm
\epsffile{\figuredir 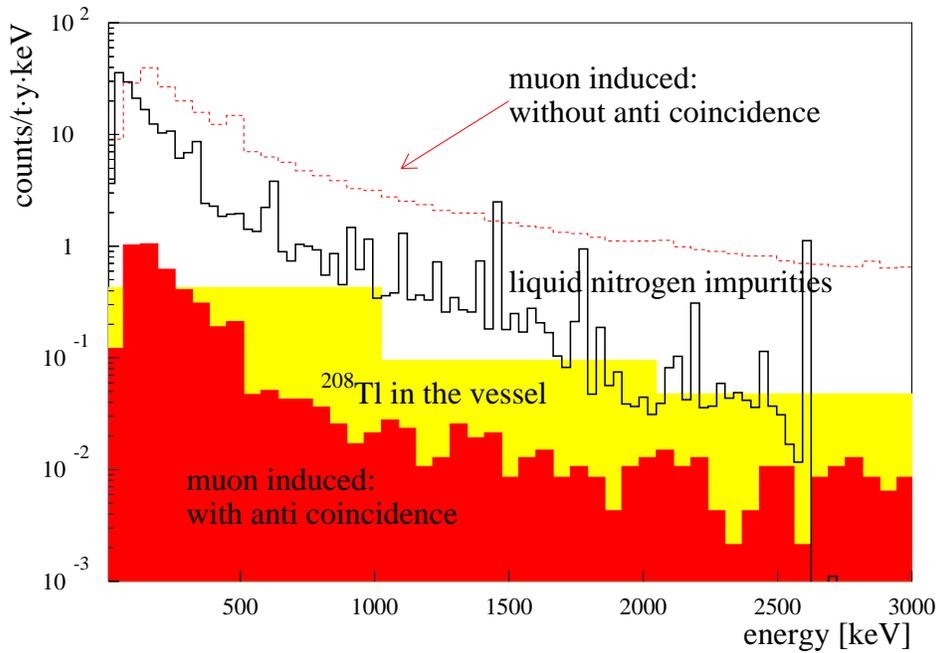}
\caption{Background from outside the nitrogen: 200 GeV muons induced events 
(dashed line) and single hit events (filled histogram); decay of
$^{208}$Tl in the steel vessel (light shaded histogram) and the background 
originating from the nitrogen impurities for comparison (thick line).}\label{fmuons}
\end{figure}

To study the influence of muons penetrating the Gran Sasso rock 
we simulated a flux of 2.3$\cdot 10^{-4}$muons/m$^{2}\cdot s$ with 200 GeV 
 \cite{arp92} crossing the tank from the top. The induced events in the germanium
detectors are shown in figure \ref{fmuons}. The
dashed histogram contains all events, whereas the filled histogram contains
only single hit events. A count rate reduction by two orders of magnitude 
through coincidence of germanium detectors from muon induced showers can be 
seen. Thus, application of this 
technique is sufficient to reduce the muon induced background far
below background from natural radioactivities.

The structure, which holds the detectors, should have, according to the
simulation of a single detector, the same purity level as the nitrogen.
The possibility to obtain organic substances with 10$^{-16}$g/g purity
levels was demonstrated by the liquid scintillator operated in CTF.

\subsection{Expected Results:}\label{chapdmlimits}
If no positive signal for \znbb{} decay will be observed, limits on the 
neutrino mass can be calculated. Assuming a radioactive background as stated
above the GENIUS experiment should measure after one year 0.8 events in the
peak region of the neutrinoless double beta decay. From this number results 
following the procedure for analysis recommended by \cite{pdg94}, which is highly conservative 
and not used in the analysis of several other $\beta\beta$ experiments, a 68 \%
upper limit on the number of \znbb{} decays of 1.2. It corresponds to a 
lower half life limit to be obtained in one year of measurement of:
\medskip
\be
T^{0\nu}_{1/2} \quad \ge \quad 5.8 \cdot 10^{27}\, y \quad \mbox{(with 68\% C.L.)} \
\ee
\medskip
Using the matrix elements of \cite{sta90} the half life limit can be converted 
into an upper limit on the neutrino mass of:
\medskip
\be
\langle m_{\nu}\rangle \quad \le 0.02 eV \quad \mbox{(with 68\%  C.L.)}\
\ee
\medskip

\begin{figure}
\epsfxsize13cm
\epsffile{\figuredir 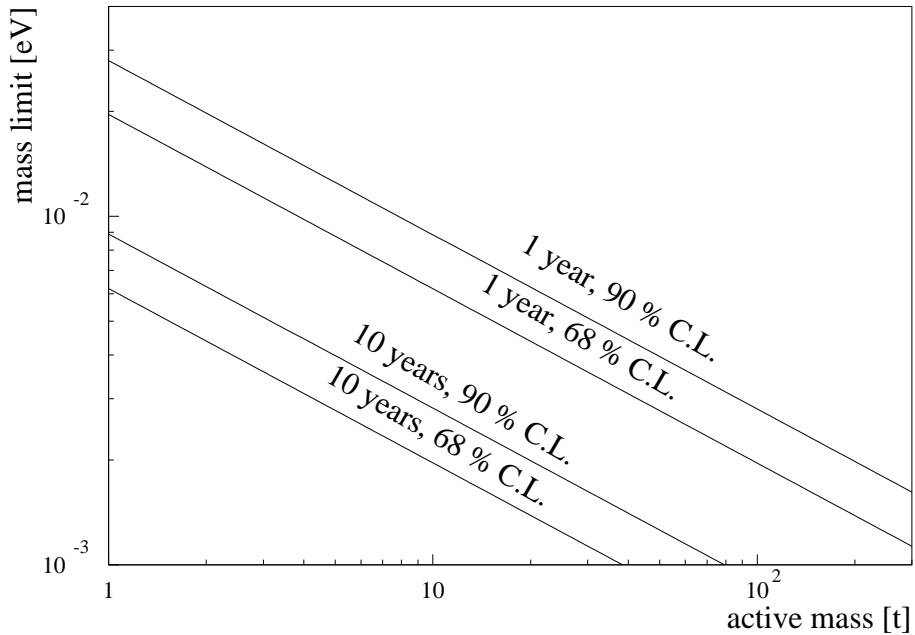}
\caption{Mass limits on Majorana neutrino mass after one and ten years measuring time as 
function of the active detector mass; zero background is assumed.}\label{fprop}
\end{figure}

\begin{figure}
\hspace*{2cm}
\epsfysize=180mm\centerline{\epsfbox{\psfiguredir 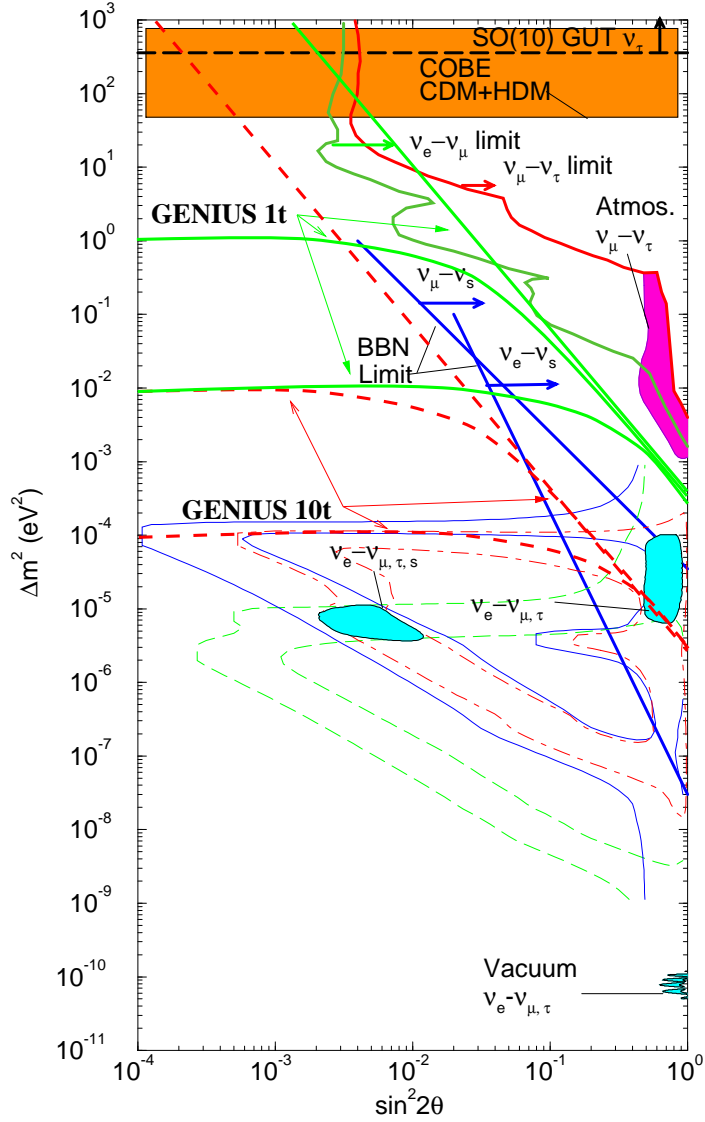}}
\caption{Ranges in the usual $\Delta$m$^{2}$ - sin$^{2}$2$\theta$
areas excluded by various neutrino oscillation experiments. Excluded
are the ranges right of the curves. The ranges {\it allowed} by
atmospheric neutrino experiments and by the { solar} experiments are
shaded. Also shown is the potential of the GENIUS experiment. The one
ton version can test the atmospheric $\nu$ problem, the 10 ton version
the large angle solution of the solar neutrino problem. The three
curves shown for each version correspond to different assumptions on
the neutrino mass hierarchy: from up to down m$_{1}$/m$_{2}$ = 0, 0.01
and 0.1({ from}  \cite{hat94} and \cite{hk97}).}\label{fhirsch}
\end{figure}

Figure \ref{fprop} shows the obtainable limits on the neutrino mass in the case
of zero background. This assumption might be justified since our assumed
impurity concentrations
 are still more conservative than proved already now for example by Borexino. 
The final sensitivity of a one ton experiment can be defined by the limit, which would
be obtained after 10 years of measurement. For the one ton experiment this 
would be:
\medskip
\be
T^{0\nu}_{1/2}\quad\ge\quad 6.4 \cdot 10^{28}\, y \quad \mbox{(with 68\% C.L.)}\
\ee
\medskip
and
\medskip
\be
\langle m_{\nu}\rangle \quad \le 0.006 eV \quad \mbox{(with 68\%  C.L.)}\
\ee
\medskip
The ultimate experiment could test { the \znbb{} half life of
$^{76}$Ge up to a limit of 5.7$\cdot$10$^{29}$y, and the }
neutrino mass down to 2$\cdot$10$^{-3}$eV using
10 tons of enriched Germanium. Such an experiment would be powerful enough to 
test neutrino oscillations on scales comparable to the best proposed,
dedicated neutrino oscillation experiments and to test the atmospheric and
the (large angle solution of the) solar neutrino puzzle (see figure 
\ref{fhirsch}). It would be stringent enough to confirm or rule out all
degenerate or inverted neutrino mass scenarios (also figure \ref{fhirsch}).
It would contribute to the search for SUSY, leptoquarks and
right--handed W mass on a sensitivity level competitive to LHC.
For a more detailed discussion we refer to \cite{hk97}.

{ With the background reached in the low--energy region GENIUS has
the potential to test the MSSM predictions for neutralinos as cold
dark matter over the whole SUSY parameter space. This is  shown in
figure \ref{zf}.}

\section{Summary and Outlook}
The GENIUS experiment using enriched $^{76}$Ge would be the most 
sensitive tool to search for neutrinoless double beta decay and cold
dark matter (WIMPs). The experiment surpasses the existing neutrino
mass experiments by a factor  of 50 to 500.
The sensitivity for WIMP nucleon scattering is four orders of magnitude
better than in the existing experiments { and at least two orders
better than} those under construction.
The shielding and purity requirements were studied using the CERN
GEANT Monte  Carlo code. Although no requirements beyond 
those of the latest solar neutrino experiments could be found, of
course further studies are needed. 
The experiment would
be small with respect to the costs compared to e.g. detectors in preparation for LHC, like
CMS or ATLAS. The proposed experiment would be a major tool of the
future - not only non--accelerator - but more general particle physics. 
It could test neutrino oscillations on a scale more sensitive than the
best proposed dedicated { accelerator} $\nu$ oscillation experiments. It could test
the atmospheric and (part of) the solar neutrino problem and would
confirm or rule out all degenerate or inverted neutrino mass
scenarios. It would contribute to the search for SUSY, leptoquarks and
left--right symmetric models on a sensitivity level competitive to
LHC. Concerning dark matter it would cover the full SUSY predicted
neutralino parameter space.
Even if SUSY would first be discovered by LHC, it would still be exciting to
test whether neutralinos will show up as dark matter (see \cite{bae97}).
For a detailed discussion of these topics we refer to a { separate} paper \cite{hk97}.

\section*{Acknowledgements}
We would like to thank Dr. G. Heusser for valuable discussions about radioactive
background requirements and Dr. J. Verplancke and Dr. P. Vermeulen of Canberra, Belgium
for good collaboration during the test experiment. We further thank
Mr. Y. Ramachers for useful discussions about dark matter.
\newpage

\bibliographystyle{unsrt}
\bibliography{/d21/klpdr/hellmig/schriften/Lib/libdiss}
\end{document}